\documentclass[conference]{IEEEtran}
\usepackage{amsmath}
\usepackage{array}
\usepackage{tabularx}
\usepackage{graphicx}
\usepackage{multirow}
\newcolumntype{Y}{>{\centering\arraybackslash}X}
\IEEEoverridecommandlockouts

\begin{document}

\title{Entropy-Assisted Multi-Modal Emotion Recognition Framework Based on Physiological Signals}

\author{
\IEEEauthorblockN{Kuan Tung$^1$$^*$\thanks{{$^*$}These two authors contributed equally.}, Po-Kang Liu$^1$$^*$, Yu-Chuan Chuang$^1$, Sheng-Hui Wang$^2$, An-Yeu (Andy) Wu$^2$}
\IEEEauthorblockA{$^1$Department of Electrical Engineering, National Taiwan University, Taipei, Taiwan}
\IEEEauthorblockA{$^2$Graduate Institute of Electronics Engineering, National Taiwan University, Taipei, Taiwan}
\IEEEauthorblockA{$^1$\{b03901039, b03901062, b03901142\}@ntu.edu.tw, $^2$\{harry, andywu\}@access.ee.ntu.edu.tw}
}

\maketitle

\begin{abstract}
As the result of the growing importance of the Human Computer Interface system, understanding human's emotion states has become a consequential ability for the computer. This paper aims to improve the performance of emotion recognition by conducting the complexity analysis of physiological signals. Based on AMIGOS dataset, we extracted several entropy-domain features such as Refined Composite Multi-Scale Entropy (RCMSE), Refined Composite Multi-Scale Permutation Entropy (RCMPE) from ECG and GSR signals, and Multivariate Multi-Scale Entropy (MMSE), Multivariate Multi-Scale Permutation Entropy (MMPE) from EEG, respectively. The statistical results show that RCMSE in GSR has a dominating performance in arousal, while RCMPE in GSR would be the excellent feature in valence. Furthermore, we selected XGBoost model to predict emotion and get 68\% accuracy in arousal and 84\% in valence.
\end{abstract}

\begin{IEEEkeywords}
Affective Computing, Multi-Scale Entropy, Multi-Scale Permutation Entropy, Extreme Gradient Boosting
\end{IEEEkeywords}

\IEEEpeerreviewmaketitle

\section{Introduction}

Due to the progression of the technology and the increasing emergence of the Internet of Things (IoT) device, building the completed human-computer interaction (HCI) system is becoming increasingly vital. However, if we hope to let the machine interacts with human more appropriately, we must give the machine an ability to consider human affect. Hence, the importance of affective computing has grown by leap and bound.

Currently, many methods have been proposed to identify people's emotional states, such as facial expression, body movements, or speech \cite{HCI1}, \cite{HCI2}. However, since human inclines to hide the true emotion inside and disguise it by the social mask, those methods mentioned above could not veridically reflect the human emotional state. In contrast, physiological measurements own several merits when developing emotion-based HCI system. Firstly, physiological signals such as EEG, ECG, and GSR cannot be easily controlled by conscious, thus the human social masking problem would be crossed out. Secondly, most physiological signals are not culturally specific, so it could be a favorable method to build an user-independent and standardized HCI system. 

Among several physiological signal datasets for emotion recognition, we selected AMIGOS \cite{AMIGOS} proposed by Juan Abdon Miranda-Correa et al. It is a dataset for Multi-modal research of human affect, which collected 40 participants’ physiological signals: Electroencephalogram (EEG), Electrocardiogram (ECG) and Galvanic Skin Response (GSR) while watching emotional videos.

\begin{figure}[t]
\centering
\includegraphics[width=8cm]{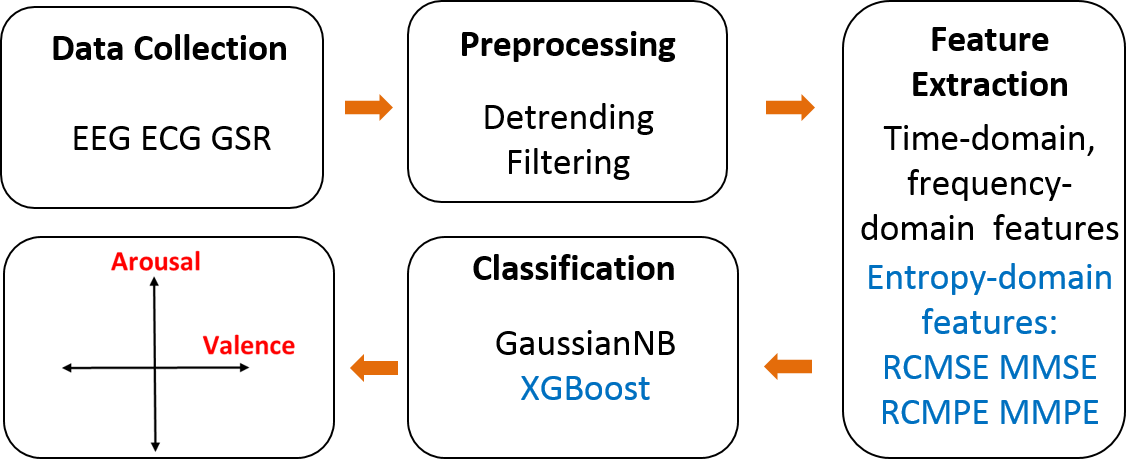}
\caption{The basic processing flow of affective computing. Black text: the original method used in AMIGOS \cite{AMIGOS}. Blue text: several modifications and improvements in our proposed framework.}
\label{figure1}
\end{figure}

AMIGOS extracted a total of 213 features in both time-domain and frequency-domain, then conducted an emotion recognition task. However, after implementing AMIGOS's emotion recognition protocol, we deduced that the features in time and frequency-domain might not be robust, thus yielded low F1-score. On the other hand, the complexity of the physiological signals has attracted a lot of notice in recent decades. Madalena Costa et al. used Multi-Scale Entropy (MSE) to separate healthy and pathologic groups successfully \cite{MSE}, Massimiliano Zanin et. al. purposed Permutation Entropy to study Epilepsy \cite{PE}. With respect to this trend, we aimed to use several entropy-based analysis to enhance the performance of emotion recognition. In this paper, we analyzed the pattern of Refined Composite Multi-Scale Entropy (RCMSE), Multivariate Multi-Scale Entropy (MMSE), Refined Composite Multi-Scale Permutation Entropy(RCMPE), and Multivariate Multi-Scale Permutation Entropy(MMPE) in different affective states. Furthermore, based on the original experimental protocol of AMIGOS, we did several modifications and implement an emotion recognition task.

Our research has two main contributions. 1) We discovered several remarkable correlations between the complexity of the physiological signals and human affect by conducting statistical analysis. 2) By applying new feature set and the different machine learning model, our classification results outperformed AMIGOS's previous results to a large extent.
 
This paper is organized into four sections as follows. First, Section \ref{sec:background} would introduce the AMIGOS dataset and the basic processing flow. Then, Our methods of extracting four kinds of entropy-domain features: RCMSE, MMSE, RCMPE, MMPE would be explained in Section \ref{sec:method}. Section \ref{sec:experiment} would list the statistical and classification results of our experiment. Finally, the conclusion and future work would be discussed in Section \ref{sec:conclusion}.

\section{Background} \label{sec:background}
\subsection{Dataset}

Up to present, several datasets have been established to let researchers undergo affective computing experiments. DEAP \cite{DEAP} is one of the most commonly used dataset in affective computing, which collected EEG, peripheral physiological signals, and face videos from the participants via clinical devices. Compared with DEAP, ASCERTAIN \cite{ASCERTAIN} is the first dataset which used commercial wearable devices and analyzed the personal traits of the participants, which is conducive to building HCI architecture. AMIGOS \cite{AMIGOS} is the newest dataset for affective computing. It recruited the participants to watch several video clips and used state-of-the-art commercial devices to collect physiological signals. It conducted self-assessment of valence, the intrinsic positive or negative feeling, and arousal, the extent of psychological excitation. Although DEAP is favorable to clinical analysis, the complicated set-up process is not favorable to building HCI system. As for ASCERTAIN, the relatively unstable and low sampling rate (32 HZ in EEG) signals might cause several problems for comprehensive analysis. In contrast, AMIGOS collected the signals in high sampling rate (128 HZ in EEG), and it also annotated other emotion states such as social context, basic emotions and external annotation of valence and arousal. Thus, it could be further applied to various types of emotion recognition tasks. In respect to these pros and cons, we selected AMIGOS to be the dataset of our experiment.

In AMIGOS, participants' emotions were annotated in range of 0 to 9 with the assessment of levels of valence and arousal. In our experiment, we split the emotional levels of into two classes: positive and negative based on the mean values of all assessment levels. This dichotomy let us could conduct the statistical analysis easier in our latter experiment.  

\subsection{Processing flow}
AMIGOS's experimental protocol is the typical processing flow of the affective computing, as shown in Fig. \ref{figure1}. First, the physiological signal from the participants would be the input of the entire flow, and it would undergo some preprocessing steps such as detrending and filtering to remove the artifact or noise. Next, several features would be extracted from the processed signals, typically from the time-domain and the frequency-domain. After then, the features would be fed into a machine learning model, and finally the machine would output its prediction of human's emotional states.

On the basis of AMIGOS's original experimental protocol, we did some modifications in order to improve the performance. In addition to time-domain and frequency-domain features, we added several entropy-domain features into our feature space, and we also replaced the classification model with XGBoost. Our modifications were marked by blue color in Fig. \ref{figure1}.
\section{Method} \label{sec:method}

\subsection{Entropy-Domain Features}
\subsubsection{Refined Composite Multi-Scale Entropy (RCMSE) \cite{RCMSE}}
Before introducing RCMSE, we first review the concept of sample entropy. It is defined as:

\begin{equation}\label{eq:se}
SampEn(x,m,r)=-\ln(\frac{n^{m+1}}{n^{m}}),
\end{equation}
where $x$ is the original time series, ${n^{m}}$ represent the total number of $m$-dimensional matched template vector pairs, $m$ is the pattern length and $r$ is the maximum tolerance. The smaller of the value of $\frac{n^{m+1}}{n^{m}}$ would lead to the higher value of SampEn, which indicates the times series is more disorder. Based on this concept, Multi-Scale Entropy(MSE) had been proposed and widely used in analyzing the physiological signal in several experiments \cite{MSE_EXP}.

RCMSE is an adaptation of MSE to resolve the problems of undefined value while doing logarithm calculation. There are two steps in RCMSE. In the first step, for each scaling factor $\tau$, $\tau$ coarse-grained time series are derived from the original time series $x$ . The $j-th$ point of $k$-th coarse-grained time series
$y_k^{(\tau)}=\{y_{k,1}^{(\tau)},y_{k,1}^{(\tau)},y_{k,2}^{(\tau)},...,y_{k,p}^{(\tau)}\}$ of $x$ is defined as follows:

\begin{equation} \label{eq:cg}
y_{k,j}^{(\tau)}=\frac{1}{\tau}\sum_{i=(j-1)\tau+k}^{j\tau+k-1}x_{i} ,\quad 1\le j\le\frac{N}{\tau},\quad 1\le k \le \tau
\end{equation},

where $x_i$ is the original time series, $N$ is the length of $x_i$.

In second step, RCMSE could be calculated as:

\begin{equation}
\begin{split}
    &RCMSE(x,\tau,m,r)=-\ln(\frac{\bar{n}_{k,\tau}^{m+1}}{\bar{n}_{k,\tau}^{m}}),
\end{split}
\end{equation}
where $\bar{n}_{k,\tau}^{m+1}=\frac{1}{\tau}\sum_{k=1}^{\tau}{n}_{k,\tau}^{m+1}\quad and    \quad\bar{n}_{k,\tau}^{m}=\frac{1}{\tau}\sum_{k=1}^{\tau}{n}_{k,\tau}^{m}$. ${n}_{k,\tau}^{m}$ represents the total number of m-dimensional matched vector pairs which is calculated from the k-th coarse-grained time series at a scale factor of $\tau$. 

Due to the length of AMIGOS dataset is at most 150 seconds, in order to avoid undefined problem when doing logarithm calculation, we set $\tau=0, 1, 2$, $m=0, 1, 2$ and $r$ = 0.2*(std of the original signal $x$) to analyze ECG, and set $\tau=1, 2, 3,...,20$, $m=2$ and $r$ = 0.2*(std of the original signal $x$) for GSR.

\subsubsection{Multivariate Multi-Scale Entropy (MMSE) \cite{MMSE}}\label{mmse}
Since there exist some correlation between different EEG channels, we use Multivariate Multi-Scale Entropy (MMSE) to analyze $p$-variate time series. The first step is to get coarse-grained time series $y_{i,j}^{\tau}$ for signals of all the considered channels which can be defined as
\begin{equation} \label{eq:mvcg}
y_{i,j}^{\tau}=\frac{1}{\tau}\sum_{t=(j-1)\tau+1}^{j\tau}x_{i,t} ,\quad 1\le i\le p,\quad 1\le j \le \frac{N}{\tau},
\end{equation}
where $i$ is the channel index and $j$ is the index of the new coarse-grained time series. Then we could create new template vector $[y_{1,j},y_{1,j+1},...y_{1,j+m_1-1},y_{2,j},,y_{2,j+1},y_{1,j+m_2-1},...,\allowbreak  y_{p,j},y_{p,j+1},...y_{p,j+m_p-1}]$. The match pairs $n^{m}$ calculation mentioned in equation \ref{eq:se} is calculated based on this new template vector, then we increased $m_i$ for times series $1$ to time series $p$ respectively in template vector and calculated the matching pairs which is denoted by  $n_{i}^{m+1}$. 
Finally $MMSE(x, \tau, r, M)$ is defined as 

\begin{equation}
MMSE(x, \tau, r, m)=-\ln(\frac{n^{m+1}}{n^{m}}),
\end{equation}
where $n^{m+1}=\frac{1}{p}\sum_{i=1}^p{n}_{i}^{m+1}$. In our experiment, we divided EEG channels into five groups: (AF3, AF4), (F7, F3, FC5, F4, F8, FC6), (T7, T8), (P7, P8), (O1, O2) based on their locations. Then, we first normalized every time series by their mean and standard deviation, and set $m_i=2$ for every $1\leq i \leq p$ and $r$ = 0.15*(std of the normalized signals) to calculated MMSE.

\subsubsection{Refined Composite Multi-Scale Permutation Entropy (RCMPE)}
Permutation entropy is another common method to evaluate the complexity of the signal \cite{PE}. For a signal of length $N$ $\{x_{1},...x_{i},...,x_{N}\}$, the PE value is defined as

\begin{equation} \label{eq:pe}
\begin{gathered}
    PE(x)=-\sum_{j=1}^{m!}{p(\pi_{j})}\log{p(\pi_{j})}, \\
    p(\pi_{j})=\frac{\#\{{i}|0\le{i}\le{i-m},(x_{i+1},...,x_{i+m})has\ type \ \pi_{j}\}}{N-m+1},
\end{gathered}
\end{equation}
where $m$ is the embedding dimension of the permutation pattern, $\{\pi_{1},...\pi_{i},...,\pi_{m!}\}$ are $m!$ distinct patterns, $j$ is the index of permutation and $p(\pi_{j})$ is the relative frequency of the permutation $\pi_{j}$. These patterns are often denoted as motifs which indicate different kinds of amplitude variation of the signals. The value of PE is always between $0$ and $\log{m!}$ where the lower bound is calculated for increasing and decreasing time series, and the upper bound for a random time series where all motifs have the same frequency. Multi-Scale PE (MPE) which incorporates coarse-graining in is often used for physiological signals due to the robust performance it brings \cite{MPE}.

RCMPE is a modified version of MPE proposed by Humeau-Heurtier \textit{et al.} \cite{RCMPE}. It overcomes the drawback of MPE where statistical reliability goes down when the coarse-graining procedure used in MPE reduces the length of the time series. There are two steps in RCMPE. First, $\tau$ coarse-grained time series $y_k^{(\tau)}=\{y_{k,1}^{(\tau)},...,y_{k,i}^{(\tau)},...,y_{k,p}^{(\tau)}\}$ are derived from the original signal $\{x_{1},...x_{i},...,x_{N}\}$ as equation (\ref{eq:cg}). Then, the RCMPE value is defined as

\begin{equation} \label{eq:rcmpe}
    RCMPE(x,\tau,m)=-\sum_{j=1}^{m!}\overline{p^{\tau}}(\pi)\ln{\overline{p^{\tau}}(\pi)},
\end{equation}
where $j$ is the index of permutation and $\overline{p^{\tau}}(\pi)$ is the average relative frequency of the permutation $\pi$ in all of the coarse-grained time series $y_k^{(\tau)}$.

In our experiment, we set $\tau=1,2,3$ and $m=2,3,4,5,6$ to analyze ECG, and set $\tau=1, 2, 3,...,20$ and $m=2,3,4,5,6$ for GSR.

\subsubsection{Multivariate Multi-Scale Permutation Entropy (MMPE)}
As mentioned in Section \ref{mmse}, we need a different approach which considers the correlation between different channels when dealing with $p$-variate time series such as EEG signals. We select MMPE proposed by Morabito \textit{et al.} \cite{MMPE}. The first step is to get coarse-grained time series $y_{i,j}^{\tau}$ for signals of all the considered channels which can be defined as shown in (\ref{eq:mvcg}). Then we calculate MMPE as

\begin{equation} \label{eq:mmpe}
    MMPE(x,\tau,m)=-\sum_{k=1}^{m!}\overline{p^{\tau}}(\pi)\ln{\overline{p^{\tau}}(\pi)},
\end{equation}
where $k$ is the index of permutation and $\overline{p^{\tau}}(\pi)$ is the average relative frequency of the permutation $\pi$ in all of the coarse-grained time series $y_{i,j}^{\tau}$.

The settings of MMPE are mostly the same with MMSE except for some differences. We set $\tau=1,2,3,...,20$ and $m=2,3,4,5,6$.

\subsection{Extreme Gradient Boosting (XGBoost)}
We select XGBoost as our classification model to predict emotion. XGBoost is a scalable and flexible machine learning method based on gradient boosting. It was proposed by Tianqi Chen and Carlos Guestrin in 2015 \cite{XGBoost}. It has become one of the most popular methods in many machine learning competitions because of the exceptional performance it shows in supervised learning problems.

The basis of XGBoost, Gradient Boosting, is an ensemble technique where a collection of predictors, commonly decision trees, are combined sequentially to become a stronger model \cite{Gradient_Boosting}. The output of the combined model can be denoted as

\begin{equation}
    \hat{y_i}=\sum_{j=1}^{T}f_j(x_i),
\end{equation}
where $f_j$ is one of the predictor, T is the total number of predictors and $x_i$ is the input feature. A specific loss function for XGBoost which is optimized at each iteration of gradient boosting is proposed as

\begin{equation} \label{eq:xgboost_loss}
    obj(\theta)=\sum_{i=1}^{n}l(y_i,\hat{y_i})+\sum_{j=1}^{T}\Omega(f_j),
\end{equation}
where $\theta$ is the parameters of the model, $l$ is the training loss function, $y_i$ and $\hat{y_i}$ are ground truth and predicted value respectively, $\Omega$ is the regularization term and $T$ is the total number of predictors. $l$ indicates how well the predictor is performing, and logistic regression is commonly used for it. $\Omega$ controls how complex the model is, and by adding it into the objective function, it can help us avoid over-fitting.

Since decision tree is typically selected as the predictor, the importance of each feature can be calculated by counting how many times a feature is used to split the data across all the trees. This can be particularly useful when evaluating the efficacy of the entropy-domain features.

\section{Experiment Setup and Results} \label{sec:experiment}
We conducted two sets of experiments: statistical analysis and classification. The first one was to discover how statistically significant the new entropy-domain features were. We could then employ the knowledge learned from it in the next step. Classification was the main task of emotion recognition since it was viewed as a binary classification problem. Our goal was to classify the classes of arousal and valence from the physiological signals of the corresponding subject.

In our experiment, only short videos were considered (There are 16 short videos per subject). The data of 7 subjects were removed due to bad signal quality and missing data in some of the modalities. Therefore, the total amount of samples in the dataset changed from 640 (40 subjects $\times$ 16 videos) to 528 (33 subjects $\times$ 16 videos).


\subsection{Statistical Analysis}
Analysis of Variance (ANOVA) was adopted for the statistical analysis of the features we extracted. It calculates the p-values by comparing the relative values between variation within groups and among groups. A common threshold for the significant statistical difference is 0.05. We used ANOVA to analyze the p-value of the entropy-domain features, as shown in Table \ref{table:ECG_RCMSE}, \ref{table:GSR_RCMSE}, \ref{table:ECG_RCMPE}, \ref{table:GSR_RCMPE}, \ref{table:EEG_MMPE}. The boldfaces are the p-values smaller than 0.1, the italicize are the p-values smaller than 0.05.

\subsubsection{RCMSE and MMSE}
The p-value of RCMSE and MMSE of ECG and GSR are shown in Table \ref{table:ECG_RCMSE} and \ref{table:GSR_RCMSE}. RCMSE of ECG has the best p-value when setting scaling factor to 2 for both arousal and valence. 
For RCMSE of GSR, we can observe that for arousal, the p-value becomes significant (0.01) when scaling factor is greater than 5. In these settings, the positive class would always have higher RCMSE, implying that the arousal of a subject is proportional to the complexity of its physiological signals. Note that the GSR would respond relatively slow according to the affect, thus the p-values become significant by increasing the scale factor. As for the p-value of MMSE of EEG, there isn't any feature whose p-value is smaller than 0.1.

\begin{table}[!t]
    \centering
    \caption{P-value of RCMSE features of ECG signals (Left:Arousal, Right:Valence)}
    \label{table:ECG_RCMSE}
    \begin{tabular}{| c | c | c | c |}
        \hline
        scale & 1 & 2 & 3 \\
        \hline
        m=0 & 0.91 &\textbf{0.08} & 0.98 \\
        \hline
        m=1 & 0.16 & \textbf{0.07} & 0.38 \\
        \hline
        m=2 & 0.13 & 0.82 & 0.11 \\
        \hline
    \end{tabular}
    \begin{tabular}{| c | c | c | c |}
        \hline
        scale & 1 & 2 & 3 \\
        \hline
        m=0 & 0.91 & \textbf{\textit{0.04}} & \textbf{0.08} \\
        \hline
        m=1 & 0.46 & \textbf{0.06} & 0.14 \\
        \hline
        m=2 & 0.76 & \textbf{\textit{0.01}} & 0.91 \\
        \hline
    \end{tabular}
\end{table}

\begin{table}[!t]
    \centering
    \caption{P-value of RCMSE features of GSR signals (A: arousal, V: valence) ($m$2 means pattern length $=$ 2)}
    \label{table:GSR_RCMSE}
    \begin{tabularx}{.48\textwidth}{|c|Y|Y|Y|Y|Y|Y|Y|Y|Y|Y|}
        \hline
            scale & 1 & 2 & 3 & 4 & 5 & 6 & 7 & 8 & 9 & 10 \\
        \hline
        A ($m$2) & 0.23 & 0.20 & 0.18 & 0.11 & 0.10 &  \textbf{0.07} & \textbf{0.05} & \textbf{\textit{0.04}} & \textbf{\textit{0.03}} & \textbf{\textit{0.03}} \\
        \hline
        V ($m$2) & 0.25 & 0.33 & 0.28 & 0.34 & 0.29 & 0.31 & 0.33 & 0.35 &	0.33 & 0.32 \\
        \hline
        scale & 11 & 12 & 13 & 14 & 15 & 16 & 17 & 18 & 19 & 20 \\
        \hline
        A ($m$2) & \textbf{\textit{0.02}} & \textbf{\textit{0.01}} & \textbf{\textit{0.01}} & \textbf{\textit{0.01}} & \textbf{\textit{0.01}} & \textbf{\textit{0.01}} & \textbf{\textit{0.01}} & \textbf{\textit{0.01}} & \textbf{\textit{0.01}} &	\textbf{\textit{0.01}} \\
        \hline
        V ($m$2) & 0.31 & 0.29 & 0.29 & 0.29 & 0.28 & 0.27 & 0.27 & 0.27 & 0.27 & 0.26 \\
        \hline
    \end{tabularx}
\end{table}

\subsubsection{RCMPE and MMPE}
The p-value of RCMPE and MMPE of ECG, GSR and EEG are shown in Table \ref{table:ECG_RCMPE}, \ref{table:GSR_RCMPE} and \ref{table:EEG_MMPE}. RCMPE of ECG performs better in valence than arousal with the significantly low p-values (0.01) in all scale. RCMPE of GSR gets greater performance in arousal since most of the features are lower than 0.05 in arousal. The positive class will have higher RCMPE which is congruent with the case in RCMSE of GSR. MMPE of EEG performs much better in valence when the scale factor goes up, which indicates the importance of the coarse-graining step.

\begin{table}[!t]
    \centering
    \caption{P-value of RCMPE features of ECG signals (A: arousal, V: valence) ($m$3 means embedding dimension $=$ 3)}
    \label{table:ECG_RCMPE}
    \begin{tabular}{|c|c|c|c|}
        \hline
        scale & 1 & 2 & 3 \\
        \hline
        A ($m$3) & \textbf{\textit{0.03}} & \textbf{\textit{0.03}} & \textbf{0.1} \\
        \hline
        V ($m$6) & \textbf{\textit{0.01}} & \textbf{\textit{0.01}} & \textbf{\textit{0.01}} \\
        \hline
    \end{tabular}
\end{table}
\begin{table}[!t]
    \centering
    \caption{P-value of RCMPE features of GSR signals (A: arousal, V: valence) ($m$2 means embedding dimension $=$ 2)}
    \label{table:GSR_RCMPE}
    \begin{tabularx}{.48\textwidth}{|c|Y|Y|Y|Y|Y|Y|Y|Y|Y|Y|}
        \hline
        scale & 1 & 2 & 3 & 4 & 5 & 6 & 7 & 8 & 9 & 10 \\
        \hline
        A ($m$2) & \textbf{\textit{0.01}} & \textbf{0.08} & \textbf{\textit{0.01}} & \textbf{\textit{0.01}} & \textbf{\textit{0.01}} & \textbf{\textit{0.01}} & \textbf{\textit{0.01}} & \textbf{\textit{0.01}} & \textbf{\textit{0.01}} & \textbf{\textit{0.01}} \\
        \hline
        V ($m$5) & \textbf{0.06} & 0.13 & 0.18 & 0.22 & 0.25 & 0.28 & 0.31 & 0.32 & 0.34 & 0.36 \\
        \hline
        scale & 11 & 12 & 13 & 14 & 15 & 16 & 17 & 18 & 19 & 20 \\
        \hline
        A ($m$2) & \textbf{\textit{0.01}} & \textbf{\textit{0.01}} & \textbf{\textit{0.01}} & \textbf{\textit{0.01}} & \textbf{\textit{0.01}} & \textbf{\textit{0.01}} & \textbf{\textit{0.01}} & \textbf{\textit{0.01}} & \textbf{\textit{0.01}} & \textbf{\textit{0.01}} \\
        \hline
        V ($m$5) & 0.37 & 0.38 & 0.39 & 0.39 & 0.40 & 0.40 & 0.41 & 0.41 & 0.41 & 0.41 \\
        \hline
    \end{tabularx}
\end{table}
\begin{table}[!t]
    \centering
    \caption{P-value of MMPE features of EEG signals (AF3, AF4) (A: arousal, V: valence) ($m$4 means embedding dimension $=$ 4)}
    \label{table:EEG_MMPE}
    \begin{tabularx}{.48\textwidth}{|c|Y|Y|Y|Y|Y|Y|Y|Y|Y|Y|}
        \hline
        scale & 1 & 2 & 3 & 4 & 5 & 6 & 7 & 8 & 9 & 10 \\
        \hline
        A ($m$4) & 0.43 & 0.46 & 0.99 & 0.82 & 0.66 & 0.45 & 0.54 & 0.52 & 0.33 & 0.47 \\
        \hline
        V ($m$6) & 0.30 & 0.37 & 0.38 & 0.46 & 0.48 & 0.46 & 0.38 & 0.30 & 0.25 & 0.15 \\
        \hline
        scale & 11 & 12 & 13 & 14 & 15 & 16 & 17 & 18 & 19 & 20 \\
        \hline
        A ($m$4) & 0.26 & 0.22 & 0.27 & 0.12 & 0.13 & \textbf{0.1} & 0.14 & 0.17 & \textbf{\textit{0.04}} & 0.17 \\
        \hline
        V ($m$4) & \textbf{0.08} & \textbf{\textit{0.03}} & \textbf{\textit{0.02}} & \textbf{\textit{0.01}} & \textbf{\textit{0.01}} & \textbf{\textit{0.01}} & \textbf{\textit{0.01}} & \textbf{\textit{0.01}} & \textbf{\textit{0.01}} & \textbf{\textit{0.01}} \\
        \hline
    \end{tabularx}
\end{table}

\subsection{Classification Results}
We used XGBoost as our classification model, fixed maximum depth and number of estimators to compensate for different sizes of input and applied grid search for parameter tuning. The classification performance was evaluated in terms of mean F1-score, which is the harmonic mean of precision and recall. The macro version of F1-score was utilized to consider both positive and negative classes. We employed leave-one-subject-out as our cross-validation scheme, where the classification models were trained using all data but videos of one subject which were then used in testing.

The emotion recognition performance is shown in Table \ref{table:cls}. Scheme I was the one reported in \cite{AMIGOS}, where 213 traditional features and Gaussian Naive Bayes were employed. Scheme II was also fed with traditional features but had XGBoost as the classification model. Scheme III was the one we proposed, which utilized entropy-domain features and XGBoost. We concatenated old traditional features with new entropy-domain features which were statistically significant (p-value $<$ 0.05). There were 41 and 101 new entropy-domain features for arousal and valence respectively.

The result shows that the Entropy-assisted model (III) we propose has the best performance in most of the situations. For single modality, all three of them in valence raise the F1-score by over or around 10\%. Huge improvements compared to previous methods (Scheme I), +12.1\% and +25.3\% for arousal and valence respectively, are found in Fusion modalities. +17.8\% can be found in EEG for valence between Scheme II and Scheme III. The aforementioned improvements prove the efficacy of the proposed new scheme. Dominant features in terms of the feature importance of Fusion modalities in Scheme III are shown in Table \ref{table:F_IMP}. entropy-domain features are highlighted in boldface. The selection of entropy-domain features (MMPE of EEG and RCMPE of ECG) vindicates the improvements in the performance of Fusion between Scheme II and III where +1.9\% and +2.9\% are found for arousal and valence.

\begin{table}[!t]
    \centering
    \caption{Mean F1-score of emotion recognition on AMIGOS \newline (up: arousal, down: valence)}
    \label{table:cls}
    \begin{tabular}{|c|c|c|c|c|}
        \hline
        Scheme (A) & EEG & ECG & GSR & Fusion \\
        \hline
        I \cite{AMIGOS} & \textbf{0.592} & 0.550 & 0.548 & 0.585 \\
        \hline
        II & 0.568 & 0.556 & 0.665 & 0.687 \\
        \hline
        III & 0.568 & \textbf{0.561} & \textbf{0.692} & \textbf{0.706} \\
        \hline
    \end{tabular}
    \begin{tabularx}{.4\textwidth}{ccccc}
      &  &  &  & \\
    \end{tabularx}
    \begin{tabular}{|c|c|c|c|c|}
        \hline
        Scheme (V) & EEG & ECG & GSR & Fusion \\
        \hline
        I \cite{AMIGOS} & 0.576 & 0.535 & 0.531 & 0.570 \\
        \hline
        II & 0.575 & 0.621 & \textbf{0.796} & 0.794 \\
        \hline
        III & \textbf{0.753} & \textbf{0.633} & \textbf{0.796} & \textbf{0.823} \\
        \hline
    \end{tabular}
\end{table}

\begin{table}[!t]
    \centering
    \caption{Dominant features in terms of the feature importance of XGBoost of Fusion modalities in Scheme III (up: arousal, down: valence, 2nd group: (F7, F3, FC5, F4, F8, FC6), 3rd group: (T7, T8), 5th group: (O1, O2))}
    \label{table:F_IMP}
    \begin{tabular}{|c|c|}
    \hline
    \multirow{11}{*}{A} & \multicolumn{1}{c|}{Dominant features} \\
    \cline{2-2}
                        & \multicolumn{1}{c|}{Spectral power of GSR in [0.0~0.2] Hz} \\
                        & \multicolumn{1}{c|}{Spectral power of ECG in [1.8~1.9] Hz} \\
                        & \multicolumn{1}{c|}{Spectral power of HRV in [0.01~0.08] Hz} \\
                        & \multicolumn{1}{c|}{Mean derivative of skin conductance slow response (SCSR)} \\
                        & \multicolumn{1}{c|}{Number of local minima in GSR} \\
                        & \multicolumn{1}{c|}{\textbf{MMPE of EEG in 3rd group ($\tau=18$, $m=2$)}} \\
                        & \multicolumn{1}{c|}{\textbf{RCMPE of ECG ($\tau=1$, $m=3$)}} \\
                        & \multicolumn{1}{c|}{Spectral power of ECG in [3.0~3.1] Hz} \\
                        & \multicolumn{1}{c|}{Spectral power of ECG in [4.3~4.4] Hz} \\
                        & \multicolumn{1}{c|}{Mean second derivative of SCSR} \\
    \hline
    \end{tabular}
    \begin{tabularx}{.4\textwidth}{ccccc}
      &  &  &  & \\
    \end{tabularx}
    \begin{tabular}{|c|c|}
    \hline
    \multirow{11}{*}{V} & \multicolumn{1}{c|}{Dominant features} \\
    \cline{2-2}
                        & \multicolumn{1}{c|}{Spectral power of GSR in [0.0~0.2] Hz} \\
                        & \multicolumn{1}{c|}{\textbf{MMPE of EEG in 2nd group ($\tau=20$, $m=6$)}} \\
                        & \multicolumn{1}{c|}{\textbf{MMPE of EEG in 2nd group ($\tau=17$, $m=6$)}} \\
                        & \multicolumn{1}{c|}{Spectral power of GSR in [0.4~0.6] Hz} \\
                        & \multicolumn{1}{c|}{\textbf{MMPE of EEG in 2nd group ($\tau=19$, $m=6$)}} \\
                        & \multicolumn{1}{c|}{Mean derivative of skin conductance slow response (SCSR)} \\
                        & \multicolumn{1}{c|}{\textbf{MMPE of EEG in 2nd group ($\tau=16$, $m=6$)}} \\
                        & \multicolumn{1}{c|}{\textbf{MMPE of EEG in 2nd group ($\tau=18$, $m=6$)}} \\
                        & \multicolumn{1}{c|}{Mean derivative of skin conductance (SC)} \\
                        & \multicolumn{1}{c|}{\textbf{MMPE of EEG in 5th group ($\tau=18$, $m=6$)}} \\
    \hline
    \end{tabular}
\end{table}

\section{Conclusion} \label{sec:conclusion}
In this paper, we propose an enhanced framework for emotion recognition. The proposed system integrates multiple entropy-domain features such as RCMSE, MMSE, RCMPE, and MMPE with XGBoost classifier. The results of statistical analysis suggest that the entropy-domain features extracted from EEG, ECG, and GSR are statistically significant for emotion recognition, especially for RCMPE of GSR in arousal. Emotion classification results show much-improved performance in classification of arousal and valence compared to previous methods.


\section*{Acknowledgment}
This work was supported by the Ministry of Science and Technology of Taiwan (MOST 106-2221-E-002-205-MY3 and MOST 106-2622-8-002-013-TA), National Taiwan University and Pixart Imaging Inc.

\bibliography{entropy}

\begin{thebibliography}{10}
\providecommand{\url}[1]{#1}
\csname url@samestyle\endcsname
\providecommand{\newblock}{\relax}
\providecommand{\bibinfo}[2]{#2}
\providecommand{\BIBentrySTDinterwordspacing}{\spaceskip=0pt\relax}
\providecommand{\BIBentryALTinterwordstretchfactor}{4}
\providecommand{\BIBentryALTinterwordspacing}{\spaceskip=\fontdimen2\font plus
\BIBentryALTinterwordstretchfactor\fontdimen3\font minus
  \fontdimen4\font\relax}
\providecommand{\BIBforeignlanguage}[2]{{%
\expandafter\ifx\csname l@#1\endcsname\relax
\typeout{** WARNING: IEEEtran.bst: No hyphenation pattern has been}%
\typeout{** loaded for the language `#1'. Using the pattern for}%
\typeout{** the default language instead.}%
\else
\language=\csname l@#1\endcsname
\fi
#2}}
\providecommand{\BIBdecl}{\relax}
\BIBdecl

\bibitem{HCI1}
Y.~L. Hsu, J.~S. Wang, W.~C. Chiang, and C.~H. Hung, ``Automatic ecg-based
  emotion recognition in music listening,'' \emph{IEEE Transactions on
  Affective Computing}, pp. 1--1, 2017.

\bibitem{HCI2}
I.~Daly, A.~Malik, J.~Weaver, F.~Hwang, S.~J. Nasuto, D.~Williams, A.~Kirke,
  and E.~Miranda, ``Identifying music-induced emotions from eeg for use in
  brain-computer music interfacing,'' in \emph{2015 International Conference on
  Affective Computing and Intelligent Interaction (ACII)}, Sept 2015, pp.
  923--929.

\bibitem{AMIGOS}
J.~Abdon Miranda-Correa, M.~Khomami~Abadi, N.~Sebe, and I.~Patras, ``Amigos: A
  dataset for mood, personality and affect research on individuals and
  groups,'' 02 2017.

\bibitem{MSE}
M.~Costa, A.~L. Goldberger, and C.-K. Peng, ``Multiscale entropy analysis of
  complex physiologic time series,'' \emph{Phys. Rev. Lett.}, vol.~89, p.
  068102, Jul 2002.

\bibitem{PE}
C.~Bandt and B.~Pompe, ``Permutation entropy: A natural complexity measure for
  time series,'' \emph{Phys. Rev. Lett.}, vol.~88, p. 174102, Apr 2002.

\bibitem{DEAP}
S.~Koelstra, C.~Muhl, M.~Soleymani, J.~S. Lee, A.~Yazdani, T.~Ebrahimi, T.~Pun,
  A.~Nijholt, and I.~Patras, ``Deap: A database for emotion analysis ;using
  physiological signals,'' \emph{IEEE Transactions on Affective Computing},
  vol.~3, no.~1, pp. 18--31, Jan 2012.

\bibitem{ASCERTAIN}
R.~Subramanian, J.~Wache, M.~K. Abadi, R.~L. Vieriu, S.~Winkler, and N.~Sebe,
  ``Ascertain: Emotion and personality recognition using commercial sensors,''
  \emph{IEEE Transactions on Affective Computing}, vol.~9, no.~2, pp. 147--160,
  April 2018.

\bibitem{RCMSE}
S.-D. Wu, C.-W. Wu, S.-G. Lin, K.-Y. Lee, and C.-K. Peng, ``Analysis of complex
  time series using refined composite multiscale entropy,'' \emph{Physics
  Letters A}, vol. 378, no.~20, pp. 1369 -- 1374, 2014.

\bibitem{MSE_EXP}
K.~Michalopoulos and N.~Bourbakis, ``Application of multiscale entropy on eeg
  signals for emotion detection,'' pp. 341--344, Feb 2017.

\bibitem{MMSE}
M.~U. Ahmed and D.~P. Mandic, ``Multivariate multiscale entropy: A tool for
  complexity analysis of multichannel data,'' \emph{Phys. Rev. E}, vol.~84, p.
  061918, Dec 2011.

\bibitem{MPE}
W.~Aziz and M.~Arif, ``Multiscale permutation entropy of physiological time
  series,'' in \emph{2005 Pakistan Section Multitopic Conference}, Dec 2005,
  pp. 1--6.

\bibitem{RCMPE}
A.~Humeau-Heurtier, C.~W. Wu, and S.~D. Wu, ``Refined composite multiscale
  permutation entropy to overcome multiscale permutation entropy length
  dependence,'' \emph{IEEE Signal Processing Letters}, vol.~22, no.~12, pp.
  2364--2367, Dec 2015.

\bibitem{MMPE}
F.~C. Morabito, D.~Labate, F.~La~Foresta, A.~Bramanti, G.~Morabito, and
  I.~Palamara, ``Multivariate multi-scale permutation entropy for complexity
  analysis of alzheimer’s disease eeg,'' \emph{Entropy}, vol.~14, no.~7, pp.
  1186--1202, 2012.

\bibitem{XGBoost}
T.~Chen and C.~Guestrin, ``Xgboost: A scalable tree boosting system,'' in
  \emph{Proceedings of the 22Nd ACM SIGKDD International Conference on
  Knowledge Discovery and Data Mining}, ser. KDD '16.\hskip 1em plus 0.5em
  minus 0.4em\relax New York, NY, USA: ACM, 2016, pp. 785--794.

\bibitem{Gradient_Boosting}
J.~H. Friedman, ``Greedy function approximation: A gradient boosting machine.''
  \emph{Ann. Statist.}, vol.~29, no.~5, pp. 1189--1232, 10 2001.

\end{thebibliography}
\bibliographystyle{IEEEtran}

\end{document}